\documentclass[twocolumn]{article}

\begin{document}

\twocolumn[
\begin{center}
{\LARGE \bf Do active galactic nuclei convert

\vspace{4pt}
dark matter into visible particles?}
\vspace{12pt}

{\Large {A. A. Grib}${\,}^{a,\, b,*}$, \ \
{Yu. V. Pavlov}${\,}^{a,\,c,\dag}$}
\vspace{12pt}

{\it ${}^a$A. Friedmann Laboratory for Theoretical Physics,
30/32 Griboedov can., St.\,Petersburg, 191023, Russia}
\vspace{4pt}

{\it ${}^b$Theoretical Physics and Astronomy
    Department, The Herzen  University, St.\,Petersburg, Russia}
\vspace{4pt}

{\it ${}^c$Institute of Mechanical Engineering,
    Russian Acad. Sci., St.\,Petersburg, Russia}
\end{center}
\hrule
\vspace{5pt}
 {\bf Abstract}
\vspace{2pt}

\hspace*{\parindent}
The hypothesis that dark matter consists of superheavy particles with
the mass close to the Grand Unification scale is investigated.
These particles were created from vacuum by the gravitation of
the expanding Universe and their decay led to the observable baryon charge.
Some part of these particles with the lifetime larger than the time of
breaking of the Grand Unification symmetry became metastable and survived
up to the modern time as dark matter.
However in active galactic nuclei due to large energies of dark matter
particles swallowed by the black hole the opposite process can occur.
Dark matter particles become interacting.
Their decay on visible particles at the Grand Unification energies leads
to the flow of ultra high energy cosmic rays observed by the Auger group.
Numerical estimates of the effect leading to the observable numbers
are given.
\vspace{7pt}

\vspace{7pt}

\vspace{4pt}
\hrule
\vspace{25pt}
]

\section{Introduction}
\footnotetext[1]{andrei\_grib@mail.ru}
\footnotetext[2]{yuri.pavlov@mail.ru}

In our papers~\cite{GrPv} the hypothesis of superheavy dark matter was
discussed.
The main argument for this hypothesis is that calculation of particle
creation from vacuum of the expanding Friedmann Universe gives finite
numerical result for the number of particles coinciding with the observable
baryon charge of the Universe (the Eddington-Dirac number) only for
particle masses close to the Grand Unification (GU) scale.
Supposing that at the GU energies these particles decayed on quarks
and leptons with breaking of the $CP$-symmetry one obtains the observable
baryon charge of the Universe.
Superheavy $X$-particles with some charge are created by gravitation in
particle-antiparticle pairs but they decay similar to $K^0$-meson decay
as short living and long living components.
It is short living component which decays on visible matter at the GU time,
long living component particles survive and if their lifetime is larger
than the time of breaking of the GU symmetry they become metastable and form
dark matter observed today.

    If similar to the $K^0$-meson theory these long living component particles
can interact with the baryon charge then one finds that some part of
$X$-particles disappeared in the early Universe and was transformed into
light particles giving the entropy of the Universe.
To get the observable density of dark matter one must have the cross-section
of such interaction $\sigma \approx 10^{-40}$\,sm${}^2$,
i.e. weak interaction.

    Recently the Auger group published their results on observation
of the ultra high energy cosmic rays (UHECR) coming from active galactic
nuclei (AGN) of Galaxies close to our Galaxy~\cite{Auger07}.
27 events with energies $E> 57 \cdot 10^{18}$\,eV were registered.

So it seems that protons of such energies can be formed at AGN.
What is the mechanism of creation of protons of such high energies at AGN?
From our point of view the natural mechanism for this effect is conversion
of superheavy dark matter into visible one near AGN.

    The AGN is formed by the supermassive black hole.
This black hole acts as a cosmic supercollider in which superheavy
particles of dark matter are accelerated close to the horizon to
the GU energies.
    For $X$-particles these energies are of the order of their mass and
such energies are quickly obtained in a short time.

The difference of our mechanism from the usual acceleration of proton
is important.
Superheavy dark matter particles colliding close to the horizon are
converted into light particles carrying the huge momenta so they can escape
the vicinity of the black hole.
The large mass of the $X$-particle is converted into energy.
Colliding protons even in case of getting a large momentum due to
the accelerating of the black hole in their back movement loose this momentum
and can not come to the Earth.

But due to our scenario of the creation of the visible matter at this
energies dark matter particles are not WIMPS (weak interacting particles)
and can be converted into visible particles.
The probability of this conversion of long living components into decaying
short living ones is defined by the nondiagonal terms of the effective
Hamiltonian for the $X, \overline{X}$ system.

The situation in the vicinity of the black hole can be different from that
in the early Universe.
The expansion of the Universe led to the decreasing of the energy of
the superheavy particles and the lifetime $\tau_l$ of the long living
component occurred to be larger than the time of breaking of the
GU symmetry.
Near black hole the time when energy is of the GU scale can be larger
than $\tau_l$.

   Now let us give some numerical estimates for the decays and interaction
of superheavy particles created by gravitation.

\section{Superheavy particles in the early Universe}
\label{SHP}

     The total number of massive particles created in
Friedmann radiation dominated Universe
(scale factor $a(t)=a_0\, t^{1/2}$)  inside the horizon is, as it
is known~\cite{GMM},
    \begin{equation}
N=n^{(s)}(t)\,a^3(t)=b^{(s)}\,M^{3/2}\,a_0^3 \ ,
\label{NbM}
\end{equation}
    where $b^{(0)} \approx 5.3 \cdot 10^{-4}$ for scalar
and  $b^{(1/2)} \approx 3.9 \cdot 10^{-3}$ for spinor particles.
It occurs that $ N \sim 10^{80} $ for $ M \sim 10^{14} $\,GeV \cite{GD}.
    The radiation dominance in the end of inflation era dark matter is
important for our calculations.
    If it is dust-like, the results will be different (see further).
    However this radiation is formed not by our visible particles.
It is quintessence or some mirror light particles not interacting with
ordinary particles.

    For the time ${t \gg M^{-1}} $ there is an era of going from the
radiation dominated model to the dust model of superheavy particles,
    \begin{equation}
t_X\approx \left(\frac{3}{64 \pi \, b^{(s)}}\right)^2
\left(\frac{M_{Pl}}{M}\right)^4 \frac{1}{M}  \,,
\end{equation}
    where $M_{Pl} \approx 1,2\cdot 10^{19}$\,GeV is Planck mass.
    If $M \sim 10^{14} $\,GeV,
$\ t_X \sim 10^{-15} $\,s for scalar and
$\ t_X \sim 10^{-17} $\,s for spinor particles.

    Let us call $t_X$ \ the ``early recombination era''.

The formula for created particles in the volume $a^3(t)$
can be written as
     \begin{equation}\label{Na}
N(t)= \left( \frac{a(t_C)}{t_C} \right)^{\! 3} b^{(s)},
\end{equation}
where $t_C=1/M$ is Compton time, $b^{(s)}$ depends on the form of $a(t)$.
From~(\ref{Na}) one can see the effect of connection of the number of
created particles with the number of causally disconnected parts on the
Friedmann Universe at the Compton time of its evolution.

    For scale factor $a(t)=a_0\, t^{\alpha}$ from Eq.~(\ref{Na})
it follows that $\ N = b^{(s)}\,M^{\, 3(1-\alpha)}\,a_0^3 $.
   Therefore for a dust-like end of inflation era one has
$ N \sim M $, and  the ratio of
the $X$-particles energy density $\varepsilon_X$ to the critical
density $\varepsilon_{crit}$  is time-independent
($\varepsilon_X < \varepsilon_{crit}$ for $M < M_{Pl}$).

   Let us define $d $, the permitted part of long-living
$X$-particles, from the condition: on the moment of
recombination $t_{rec} $ in the observable Universe one has
$
d\,\varepsilon_X(t_{rec}) =\varepsilon_{crit}(t_{rec})  \,.
$
    It leads to
\begin{equation}
d=\frac{3}{64 \pi \, b^{(s)}}\left(\frac{M_{Pl}}{M}\right)^2
\frac{1}{\sqrt{M\,t_{rec}}}\, .
\label{d}
\end{equation}
     For $M=10^{13} - 10^{14} $\,GeV one has
$d \approx 10^{-12} - 10^{-14} $ for scalar and
$d \approx 10^{-13} - 10^{-15} $ for spinor particles.
    So the lifetime of the main part or all $X$-particles must be smaller
or equal than $t_X$.

     Now let us construct the model which can give: \
{\bf (a)} short-living $X$-particles decay in time
   $\tau_q < t_X $ (more wishful is
   $\tau_q \sim t_C \approx 10^{-38} - 10^{-35} $\,s,
i.e., the Compton time for $X$-particles); \
{\bf (b)} long-living particles decay with $\tau_l > 1/M $.
   Baryon charge nonconservation with $CP$-nonconservation in full
analogy with the $K^0$-meson theory with nonconserved hypercharge and
$CP$-nonconservation leads to the effective Hamiltonian of the decaying
$X, \bar{X}$ - particles with nonhermitean matrix.

    For the matrix of the effective Hamiltonian
$ H=\{ H_{ij} \}, \ {i,j=1,2}$  let $H_{11}\! =\! H_{22}$
due to $CPT$-invariance.
    Denote
$\ \varepsilon=(\sqrt{\vphantom{ }H_{12}} - \sqrt{H_{21}}\,)\, / \,
(\sqrt{H_{12}} + \sqrt{H_{21}} \, )$.
   The eigenvalues $\lambda_{1,2} $ and eigenvectors
$|\Psi_{1,2}\rangle $  of matrix $H$ are
    \begin{equation}
\lambda_{1,2} = H_{11} \pm \frac{H_{12}+H_{21}}{2} \,
\frac{1-\varepsilon^2}{1+\varepsilon^2} \,,
\end{equation}
    \begin{equation}
|\Psi_{1,2}\rangle =\frac{1}{\sqrt{2\,(1+|\varepsilon |^2)}}\,
\left[ (1+\varepsilon) \,|1\rangle \pm \,(1- \varepsilon) \,
 |2\rangle \right].
\end{equation}
    Let us choose matrix of effective Hamiltonian as
\begin{eqnarray}
\!\!\!\! H \!\!=\!\!\!     \left(
\begin{array}{cc}   \!\!\!\!\!\!\!\!\!\!
E-\frac{i}{4}(\tau_q^{-1} \!+\tau_l^{-1})
  &                 \!\!\!\!\!\!\!\!
\frac{1+\varepsilon}{1-\varepsilon}
[A \!-\frac{i}{4}(\tau_q^{-1} \!-\tau_l^{-1})]    \!\!
 \\  & \\           \!\!\!
\frac{1-\varepsilon}{1+\varepsilon}
[A \!-\frac{i}{4}(\tau_q^{-1} \!-\tau_l^{-1})]
 &                  \!\!\!
E-\frac{i}{4}(\tau_q^{-1} \!+\tau_l^{-1})      \!\!\!\!\!\!   \\
\end{array}        \right)\!\!.  \hspace{-17pt} \nonumber \\
\label{HM}
\end{eqnarray}
     Then the state $|\Psi_1 \rangle $ describes
short-living particles $X_q$ with the lifetime
$ \ \tau_q \ $ and mass $E+A$.
    The state $\ |\Psi_2 \rangle $ is the state of long-living particles
$X_l$ with lifetime $ \tau_l \ $  and mass $E-A$.
Here $A$ is the arbitrary parameter $-E<A<E$  and it can be zero, then $E=M$.

     If $\tau_l$ is larger than the time of breaking of the Grand Unification
symmetry it can be that the some quantum number can be conserved leading to
some effective time
 $\tau_l^{\rm eff} > t_U \approx 4.3 \cdot 10^{17}$\,s
\  ($t_U $ is the age of the Universe).
    The small $ d \sim 10^{-15} - 10^{-12} $ part of long-living
$X$-particles with $\tau_l > t_U $ forms the dark matter.

For $ t_l^{\rm eff} \le 10^{27}$\,s one could have the observable flow of
UHECR from the decay in our Galaxy~\cite{BBV}.
But in this case one must get a strong anisotropy in the direction to
the center of the Galaxy~\cite{ABerK}.
However, Auger experiments don't show such an anisotropy and one must
suppose $ t_l^{\rm eff} > 10^{27}$\,s.

    Use a model with effective Hamiltonian~(\ref{HM}),
where $\tau_l > t_U$, and now take into account vanishing of long-living
component with conversion to light particles due to interaction with
baryon matter.
    Consider a model with an interaction which in the
basis   $\ |1 \rangle, \ |2 \rangle $  is described by the matrix
     \begin{equation}
H^d =     \left(
\begin{array}{cc}
0  & 0   \\
0  & - i \gamma \\
\end{array}        \right).
\label{Hd}
\end{equation}
     The eigenvalues of the Hamiltonian  $H+H^d$  are
     \begin{eqnarray}
\lambda^d_{1,2} \! &=& \! E - \frac{i}{4}
\left(\tau_q^{-1} + \tau_l^{-1} \right) - i\,\frac{\gamma}{2} \, \pm
\nonumber  \\
&\pm& \! \sqrt{ \left( A - \frac{i}{4} \left(\tau_q^{-1} - \tau_l^{-1}
\right) \right)^2 -\frac{\gamma^2}{4} } \ .
\label{lamdop}
\end{eqnarray}
    In case $\gamma \ll \tau_q^{-1}$,
for the long-living component one obtains
     \begin{equation}
\lambda^d_{2} \approx  E - A - \frac{i}{2}\, \tau_l^{-1}
-i\,\frac{\gamma}{2} \,,
\label{ldolg}
\end{equation}
     \begin{equation}
\| \Psi_2(t) \|{}^2 = \| \Psi_2(t_0) \|{}^2 \exp \left[
\frac{t_0 - t}{\tau_l} - \int_{t_0}^t \! \gamma(t)\, d t \right]\!.
\label{P21}
\end{equation}

The parameter $\gamma$,  describing the interaction with the
substance of the baryon medium, is evidently dependent on its state
and concentration of particles in it.
   For approximate evaluations take this parameter as
proportional to the concentration of particles:
$\gamma = \alpha\, n^{(s)}(t)$.
       For
$\tau_l \gg t_U$, \ $t \le t_U$, \ $a(t)=a_0 \sqrt{t}$,
by Eq.~(\ref{NbM}),  one obtains
     \begin{equation}
\!\!\! \| \Psi_2(t) \|^2 \!=\! \| \Psi_2(t_0) \|^2 \exp \biggl[
\alpha 2 b^{(s)} \! M^{3/2} \! \biggl( \! \frac{1}{\sqrt{t}} -
\frac{1}{\sqrt{t_0}} \! \biggr) \! \biggr].
\label{Ptt0}
\end{equation}
    So the decay of the long-living component due to this mechanism
takes place close to the time $t_0$.

    One can think that this interaction of $X_l$ with baryon charge
is effective for times, when the baryon charge becomes strictly
conserved, i.e., we take the time larger or equal to the electroweak
time scale, defined by the temperature of the products of decay
of $X_q$.
    Supposing the great difference in masses of $X_q$ and the products of
its decay ("the great desert") one can assume the products as
ultra relativistic gas with nonzero entropy.
    Its temperature is defined from
$ M n^{(s)}(\tau_q) \approx \sigma T^4 $
and is given by
     \begin{equation}
T(t) = \left( \frac{ 30\, b^{(s)} }{ \pi^2 N_l } \right)^{\!1/4}
\frac{ M^{5/8}\, \tau_q^{1/8} }{ k_B\, \sqrt{t} }\,,
\label{T}
\end{equation}
   where $k_B$ is Boltzmann constant, and $N_l$ is defined by the number
of boson $N_B$ and fermion $N_F$ degrees of freedom of all kinds of
light particles:
$N_l=N_B + \frac{7}{8} N_F$\, (see Ref.~\cite{KKZ}).
    At time $t_X$, this temperature is equal to
     \begin{equation}
T(t_X) = \frac{64 \sqrt{\pi}}{3}\!\!
\left( \frac{ 30}{N_l} \right)^{\!1/4}\!\!\!
\left( b^{(s)} \right)^{\!5/4}\!\!
\left( M \tau_q \right)^{\!1/8}\!\!  \frac{ M^3}{k_B M_{Pl}^2}.
\label{TtX}
\end{equation}
    If $ \tau_q = 1/M $ and $N_l\sim 10^2$ -- $10^4 $, then for spinor
$X$-particles $ T(t_X) \approx 300$ -- $100 $\,GeV,
i.e., the electroweak scale for created particles
(which is however different from that for the background).
This unexpected coincidence shows consistency of our reasonings.

    So let us suppose $t_0 \approx t_X$.
    If  $d$  is the part of long-living particles surviving up to
the time  $t$   $\, (t_U \ge t \gg t_C)$,  then from Eqs.~(\ref{d})
and (\ref{Ptt0}) one obtains the evaluation for the parameter~$\alpha$:
     \begin{equation}
\alpha = \frac{ - 3 \ln d}{ 128 \pi (b^{(s)})^2} \,
\frac{M_{Pl}^2}{M^4} \,.
\label{ald}
\end{equation}
    For  $M=10^{14}$\,GeV  and  $d=10^{-14}$ one obtains
$\alpha \approx 10^{-40}$\,cm${}^2$.
    If  $\tau_q \sim 10^{-38} - 10^{-35} $\,s
then the condition  $\gamma(t) \ll \tau_q^{-1} $
used in Eq.~(\ref{ldolg}) is valid for $t> t_X$.
    Thus such a mechanism of the decay of the long-living component of
$X$-particles was important in the early Universe
at $t_0 \approx t_X$.

The observed entropy in this scenario originates due to transformation of
$X$-particles into light particles: quarks, antiquarks and some particle
similar to $\Lambda^0$ in $K^0$-meson theory, having the same quantum
number as $X$.
Baryon charge is created close to the time $t_q $, which can be equal to
the Compton time of $X$-particles $t_C \sim 10^{-38} - 10^{-35} $\,s.

    Our scheme can also work for spinor particles.
Then it is possible to investigate some version of the {\it see-saw\/}
mechanism~\cite{GMRS} for Majorana neutrinos in the Grand Unification
theory, so that heavy sterile neutrinos form the dark matter.

\section{Superheavy particles as source UHECR from active galaxy nuclei}

Now let us give some numerical estimates of the conversion of superheavy
particles into UHECR in active galactic nuclei.
The Auger group registered 27 UHECR with energies higher than
$57 \cdot 10^{18}$\,eV.
The integrated exposure of Auger observatory for these data is
$9.0 \times 10^3$\,km${}^2\,$sr\,year.
The Auger group found the correlation of UHECR with nearby active
extragalactic objects~\cite{Auger07}.
There are 318 AGN on the distance smaller than 75\,Mpc.
It is easy to see that if these AGN are distributed uniformly and have
the same intensity of UHECR radiation, each of the AGN must radiate
approximately $j=10^{39}$ UHECR in a year.
    The distance of propagation of UHECR is limited by
the Greizen-Zatsepin-Kuzmin limit~\cite{GZK} and for proton with the energy
higher than $8\cdot 10^{19}$\,eV this distance cannot be larger than 90\,Mpc.

    Due to Auger results the source of 2 particles of superheavy energy
is the AGN Centaurus A located on 11 million of light years from the Earth.
It is easy to calculate that for the integral exposure of Auger observatory
this AGN must radiate approximately  $3\cdot 10^{37}$ UHECR in a year.

Our hypothesis is that these UHECR in AGN arise due to superheavy dark
matter particles converted into quarks and leptons at high energies obtained
by them close to the supermassive black hole horizon of the AGN.
    To understand the process of this conversion one must remember
the consideration  of parts~\ref{SHP} of this paper.

Superheavy dark matter particles with mass $M=10^{14}$\,GeV, fall on the
black hole, so that if 100 \% of these particles are converted into UHECR
than it's mass must have the order of $10^{28}$\,g.
    Even if only $\eta = 10^{-4}$ of the total mass of superheavy particles
close to the horizon is converted into ordinary particles the whole mass
of dark matter $ m=M j/\eta $ is much lower than the mass of the ordinary
matter accreted on the black hole leading to it's observed light radiation.

    The source of the energy of UHECR is the decay of superheavy particles
at the GU energies on quarks and leptons which due to our reasoning at
part~\ref{SHP} of our paper led to the origination of the baryon charge
of the Universe.
    The mass of superheavy particle is converted into the energy of light
particles the flow of which can go from the black hole to the Earth
similar to photons.
    The black hole plays the role of the cosmic accelerator or
super collider creating the conditions for transforming the long living
component of $X$-particles into short living and it's decay.
In any case if the time of existence of the $X$-particles with GU energies
differently from the situation at the early Universe is larger
than $\tau_l$,\ $X_l$ must decay.

    Now let us evaluate the density of dark matter needed to form
the observable flow of UHECR.
    Suppose that dark matter is uniformly distributed with the density typical
for the ordinary matter in central parts of the galaxies
$\rho = 10^{-20}$\,g/sm${}^3$.
Let us take the typical velocities of the dark matter particles in large
distance from the central black hole as $v_\infty \approx 10^8$\,sm/s
(these are star velocities in the central parts of galaxies).
The capture cross-section of the non-relativistic particles by
a Schwarzschild black hole
is given by (see Eq.(3.9.1) in~\cite{ZelNovTTES})
     \begin{equation}
\sigma_{c} = 4 \pi \left( \frac{c}{v_\infty} \right)^2 r_g^2 \,.
\label{sz}
\end{equation}
Here $r_g$ is the horizon radius of the black hole.
Taking $M_{BH} = 10^8 M_S $ one obtains for the velocity of the accretion
of the dark matter on the black hole
$v_a = \Delta M / \Delta t =
\sigma_c v_\infty \rho \approx 3 \cdot 10^{28}$\,g/year
which is consistent with our evaluation of the Auger observation
$j=10^{38}$ UHECR/year.

    One must mention that conversion of dark matter into UHECR is effective
only for objects with large quantity of the diffusive dark matter close to
the black hole.
This situation can occur only for AGN and is improbable for ordinary galaxies.
From~(\ref{sz}) one can see that capture of dark matter by the black hole
is proportional to the square of black hole mass, so that the flow of UHECR
from black hole of star masses is negligible.
    We don't have observation data for the distribution of dark matter
at central regions of galaxies with AGN.
If one takes for the distribution density of dark matter the numerical
profiles
     \begin{equation}
\rho(r) = \frac{\rho_0}{(r/r_0)^\beta (1+r/r_0)^{3-\beta}}
\label{rNFW}
\end{equation}
with $\beta=1$ for Navarro-Frenk-White profile~\cite{NFW96},
$\beta=1.5$ for Moore profiles, $r_0=45$\,kpc,
$\rho_0 =10^{-24}$\,g/sm${}^3$ \cite{ABerK},
then one again obtains
$ v_a \sim 2 \cdot 10^{28} - 10^{30}$\,g/year.

So our evaluation leads to reasonable quantity of the accretion of
supermassive dark matter particles on the black hole.
This dark matter can be considered to be a source  of UHECR arising
from the decay of supermassive particles on visible matter close
to the horizon of the supermassive black hole.

    Now let us discuss the possible physical mechanism of conversion of
dark matter into visible matter at AGN.
It is reasonable to think that AGN differently from other black holes
are rapidly rotating supermassive black holes.
Then one has the well known Penrose mechanism~\cite{Penrose69}.
    The incoming particle in ergosphere decays on two particles,
one with negative energy goes inside the black hole while another
particle with the opposite momentum and the energy larger than the incoming
one goes to the outside space.
The condition for the conversion of dark matter superheavy particles into
quarks and leptons is great relative energy-momentum in interaction of
these particles.
This condition can be fulfilled for our Penrose process.

    Then the particle with the energy greater than the GU scale going in
opposite direction to AGN can collide with the other superheavy particle
falling inside and so on.
In the result macroscopic amount of dark matter can be "burned" close
to the AGN.
So AGN can work as a great cosmical collider.

\vspace{4mm}
\noindent
{\bf Acknowledgements}.

\vspace{2mm}
This work was supported by the Ministry of Science and Education
of Russia, grant RNP.2.1.1.6826.


\end{document}